\documentclass[english,smallextended]{svjour3}
\usepackage[T1]{fontenc}
\usepackage[latin9]{inputenc}
\usepackage{verbatim}
\usepackage{url}
\usepackage{amsmath}
\usepackage{amssymb}
\usepackage{graphicx}
\usepackage{esint}

\makeatletter

\RequirePackage{fix-cm}

\smartqed  

\usepackage{hyperref}

\makeatother

\usepackage{babel}
\begin{document}

\title{Testing Chern-Simons modified gravity with orbiting superconductive
gravity gradiometers \\
--- The non-dynamical formulation}

\titlerunning{Testing CS gravity with orbiting SGG}

\author{Li-E Qiang \and Peng Xu}

\authorrunning{L. Qiang and P. Xu}

\institute{Li-E Qiang \at  Department of Geophysics, College of the Geology
Engineering and Geomatics, \\
Chang'an University, Xi'an, 710054, China.\\
\email{qqllee815@chd.edu.cn}\\
\and  Peng Xu \at  Academy of Mathematics and Systems Science, Chinese
Academy of Sciences, \\
No.55, Zhongguancun Donglu Street, Beijing, 100190, China.\\
Fax: 86-10-62541689\\
\email{xupeng@amss.ac.cn}}
\maketitle
\begin{abstract}
High precision Superconductivity Gravity Gradiometers (SGG) are powerful
tools for relativistic experiments. In this paper, we work out the
tidal signals in non-dynamical Chern-Simons modified gravity, which
could be measured by orbiting SGGs around Earth. We find that, with
proper orientations of multi-axes SGGs, the tidal signals from the
Chern-Simons modification can be isolated in the combined data of
different axes. Furthermore, for three-axes SGGs, such combined data
is the trace of the total tidal matrix, which is invariant under the
rotations of SGG axes and thus free from axis pointing errors. Following
nearly circular orbits, the tests of the parity-violating Chern-Simons
modification and the measurements of the gravitomagnetic sector in
parity-conserving metric theories can be carried out independently
in the same time. A first step analysis on noise sources is also included.\\

\end{abstract}

\section{Introduction\label{sec:Introduction} }

Among the modified gravitational theories, the extensions to the Einstein-Hilbert
action with second order curvature terms are of particular interest,
which may arise in the full, but still lacking, quantum theory of
gravity as high energy corrections to GR, see \cite{Niedermaier2006}.
The string theory inspired Chern-Simons (CS) modified gravity 
\cite{Deser1982,Campbell1990,Campbell1991,Jackiw2003,Alexander2009},
with the additions of a parity-violating term $R\;^{\star}R$ and
a coupling scalar field $\theta$, is one of such extensions of GR.
Being a promising model, CS modified gravity has found connections
with different fields such as gravitational physics, particle physics,
string theory, loop quantum gravity, and cosmology, please consult
\cite{Alexander2009} for detailed discussions.

CS modified gravity now contains two classes of formulations, the
non-dynamical and dynamical formulations, which are in fact two distinct
theories. In the non-dynamical formulation, the CS scalar $\theta$
is externally prescribed. The so called canonical choice is to set
$\theta$ as a linear function of coordinate time proportional inversely
to a mass scale $M_{CS}$ \cite{Jackiw2003}. While, in the more realistic
but complicate dynamical formulation, the evolution of the CS scalar
is then sourced by the spacetime curvature. The non-dynamical formulation
serves now as a useful model that provides us insights into parity-violating
theories of gravity. Up to now, the experimental tests and constraints
on CS gravity are all based on observations from astrophysics and
space based experiments. The first but weak bound on the canonical
CS scalar $\theta$ was obtained in \cite{Smith2008} based on the
results from LAGEOS I, II \cite{Ciufolini2004,Ciufolini2010a} and
Gravity Probe-B \cite{Everitt2011} missions, which had constrained
the CS mass scale to $M_{CS}\gtrsim2\times10^{-13}eV$. The strongest
bounds on the canonical scalar up to now was based on the data from
double binary pulsars \cite{Yunes2009a}, which had the constraint
$M_{CS}\gtrsim4.7\times10^{-10}eV$ as been revised in \cite{Ali-Haimoud2011a}.
For dynamical CS gravity, the vacuum solutions outside the rotating
black holes and stars were studied in the slow rotation approximation
in \cite{Yunes2009b,Ali-Haimoud2011,Yagi2012a}, and their possible
tests can be found in \cite{Chen2010,Yagi2013,Vincent2013}. Moreover,
the parity-violating term $R\;^{\star}R$ also leaves distinguishable
signatures in gravitational waves, which may be captured by ground
based or future space borne gravitational wave antennas, please consult
\cite{Sopuerta2009,Garfinkle2010,Pani2011,Canizares2012} for details.

In experimental relativity, Braginskii and Polnarev had obtained for
relativistic gravitational theories an interesting spin-quadrupole
coupling between rotating sources and orbiting quadrupole oscillators
\cite{Braginskii1980,Polnarev1986}. Along this line, the principles
of detecting Earth gravitomagnetic field with orbiting gradiometers
are studied in a series of works of Paik, Mashhoon and their collaborators
\cite{Mashhoon1982,Theiss1985,Paik1988,Paik1989,Mashhoon1989,Paik2008}.
Today, gradiometers have already been employed in space based experiment
such as the one on GOCE satellite \cite{Rummel2011}. The high precision
Superconductive Gravity Gradiometer (SGG) have been developed at the
University of Maryland by Moody, Paik and their colleagues, and their
performance level has already reached $2\times10^{-11}s^{-2}/Hz^{\frac{1}{2}}$
in 1990s \cite{Moody2002}. Recently, with the improvement in cryogenics
and the magnetically levitated test masses, a new design scheme with
$3$ orders of magnitude improvement in the SGG performance over the
frequency band $0.5mHz\sim0.1Hz$ has been developed by this group
\cite{Moody2010,Shirron2010}. It is also pointed out in \cite{Paik2006}
that a performance level of $10^{-15}s^{-2}/Hz^{\frac{1}{2}}$ over
this band is within the capability of the SGGs under development.
One can consult \cite{Paik1989,Paik2008,Li2014} for the detailed
error analysis of the relativistic gradient measurements with SGGs
in space. On the other hand, as pointed out by Alexander, Yunes and
etc. \cite{Alexander2007,Alexander2007a} that, within the weak field
and slow motion limits, the non-dynamical CS gravity differs from
General Relativity (GR) only in the gravitomagnetic sector. Based
on these results, we study in this work the theoretical principles
of testing the parity-violating non-dynamical CS gravity with SGGs
in space.

\section{Signatures of gradient measurements in the non-dynamical Chern-Simons
gravity\label{sec:Signatures-of-gradient}}

This work is heavily based on the mission concepts studied in \cite{Paik1988,Paik1989,Mashhoon1989},
that the gradient force from Earth gravitomagnetic field are measured
by orbiting SGGs along nearly circular orbits. We worked out in this
section the signatures of the gradient observable in non-dynamical
CS gravity.

\subsection{The Non-dynamical Chern-Simons modified gravity\label{sub:The-Non-dynamical-Chern-Simons}}

We first give a brief introduction to the non-dynamical formulation
of CS modified gravity, for detailed discussions please consult \cite{Alexander2009,Alexander2007,Alexander2007a}.
The geometric units $c=G=1$ are adopted in the followings. The action
for non-dynamical CS gravity reads
\[
S:=S_{GR}+S_{CS}+S_{matt},
\]
where
\begin{eqnarray}
S_{GR} & = & \frac{1}{16\pi}\int d^{4}x\sqrt{-g}R,\quad\: S_{CS}=\frac{\alpha}{4}\int d^{4}x\sqrt{-g}\theta R^{\star}R,\label{CSaction}
\end{eqnarray}
and $S_{matt}$ is the action from the matter fields that is independent
of $\theta$. $g$ is the determinant of the metric and the Pontryagin
density
\begin{equation}
R\;^{\star}R=\frac{1}{2}\epsilon^{cdef}R_{\; bef}^{a}R_{\; acd}^{b}.\label{eq:pon}
\end{equation}
The CS coupling field $\theta$ is externally prescribed and depends
on the specific theory that under consideration. $\theta$ can be
viewed as the deformation function, and the difference between CS
gravity and GR is proportional to the deformation parameters $\nabla_{a}\theta$
and $\nabla_{a}\nabla_{b}\theta$. In the so called canonical CS coupling
$\theta$ is a spatially isotropic function and depends linearly on
the coordinate time $t$ \cite{Jackiw2003}, therefore the deformation
parameter contains only $\dot{\theta}$.

The field equation of the non-dynamical CS gravity is obtained by
varying the action with respect to the metric
\begin{equation}
R_{ab}-\frac{1}{2}g_{ab}R+16\pi\alpha C_{ab}=8\pi T_{ab},\label{eq:field_eq}
\end{equation}
where $C_{ab}$ is the 4-dimensional generalization of the Cotton-York
tensor
\begin{equation}
C^{ab}=\nabla_{c}\theta\epsilon^{cde(a}\nabla_{e}R_{\; d}^{b)}+\frac{1}{2}\nabla_{c}\nabla_{d}\theta\epsilon^{efd(a}R_{\;\;\; fe}^{b)c}.\label{eq:cotton}
\end{equation}
The introduction of the new scalar degree of freedom $\theta$ also
gives rise to the new constraint
\begin{equation}
\nabla_{a}C^{ab}=-\frac{1}{8}\nabla^{b}\theta(^{\star}RR)=0.\label{eq:constrain}
\end{equation}
If the above constraint is satisfied, from Eq.(\ref{eq:field_eq})
the Bianchi identities and the equations of motion for matter fields
$\nabla_{a}T^{ab}=0$ are recovered, which rank the non-dynamical
CS gravity a metric theory.

In the weak field and slow motion limits, the Parametrized Post-Newtonian
(PPN) metric of the non-dynamical CS gravity outside a compact source
was carefully worked out in \cite{Alexander2007,Alexander2007a}.
As mentioned before, the non-dynamical CS gravity differs from GR
only in the gravitomagnetic sector
\begin{equation}
g_{0i}^{CS}=g_{0i}^{GR}+\chi(r\nabla\times\mathbf{V})_{i},\label{eq:CSGM}
\end{equation}
here $r$ denotes the distance to the mass center of the compact source
and $V_{i}$ is the PN potential, see Appendix.\ref{sec:The-standard-PPN}.
The dimensionless parameter $\chi=\frac{32\pi\alpha\dot{\theta}}{r}$
is the new PN parameter for non-dynamical CS gravity, and the CS mass
scale is defined as \cite{Alexander2007,Alexander2007a}
\begin{equation}
Mcs\equiv\frac{1}{8\pi\alpha\dot{\theta}}=\frac{4}{\chi r}.\label{eq:mass}
\end{equation}

\subsection{The basic settings\label{sub:The-basic-settings}}

We restrict ourselves to the so-called ``semi-conservative'' metric
theories, which are based on action principles and respect the conservation
law of 4-momentum \cite{Will2014}. Therefore, we work with the metric
theories with four relevant standard PPN parameters $\{\gamma,\ \beta,\ \xi,\ \alpha_{1},\ \alpha_{2}\}$
together with the additional CS parameter $\chi$, please see \cite{Will2014}
or Appendix.\ref{sec:The-standard-PPN} for the PPN formalism. The
PN coordinates system $\{t,x^{i}\}$ outside Earth is chosen as follows.
The mass center of Earth is set at the origin. The basis $(\frac{\partial}{\partial x^{3}})^{a}$
is set to parallel to the direction of the Earth angular momentum
$\mathbf{J}$, $(\frac{\partial}{\partial x^{1}})^{a}$ is pointing
to a reference star $\Upsilon$ and $(\frac{\partial}{\partial x^{2}})^{a}$
determined by the right-hand rule $(\frac{\partial}{\partial x^{1}})^{a}\times(\frac{\partial}{\partial x^{2}})^{a}=(\frac{\partial}{\partial x^{3}})^{a}$.
Such coordinate directions are tied to the remote stars, and the time
$t$ is measured by the observers at asymptotically flat region. Within
our coordinate system the PN metric outside Earth reads
\begin{eqnarray*}
g_{00} & = & -1+2U-2\beta U^{2}-2\xi\Phi_{W}+(2\gamma+2-2\xi)\Phi_{1}\\
 &  & +2(3\gamma-2\beta+1+\xi)\Phi_{2}+2\Phi_{3}+2(3\gamma-2\xi)\Phi_{4}\\
 &  & +2\xi\mathcal{A}-(\alpha_{1}-\alpha_{2})w^{2}U-\alpha_{2}w^{i}w^{j}U_{ij}-2\alpha_{1}w^{i}V_{i}+\mathcal{O}(\epsilon^{6}),\\
g_{0i} & = & -\frac{1}{2}(4\gamma+3+\alpha_{1}-\alpha_{2}-2\xi)V_{i}-\frac{1}{2}(1+\alpha_{2}+2\xi)W_{i}\\
 &  & +\chi r(\nabla\times\mathbf{V})^{i}-\frac{1}{2}(\alpha_{1}-2\alpha_{2})w^{i}U-\alpha_{2}w^{j}U_{ij}+\mathcal{O}(\epsilon^{5}),\\
g_{ij} & = & (1+2\gamma U)\delta_{ij}+\mathcal{O}(\epsilon^{4}),
\end{eqnarray*}
please see Appendix.\ref{sec:The-standard-PPN} for the PN potentials.
For low and medium Earth orbits experiments, the magnitude of $\epsilon$
is about $10^{-5}.$

We model Earth as an ideal and uniform rotating spherical body, and
for the effects from Earth gravity multiples on the gravitomeganetic
field measurements with orbiting SGGs, please consult \cite{Li2014}.
The preferred-frame and the preferred-location effects are tightly
constrained by observations, and we now have the upper bounds of the
related PN parameters as $\alpha_{1}\sim4\times10^{-5},\ \alpha_{2}\sim2\times10^{-9},\ \alpha_{3}\sim4\times10^{-20}$
and $\xi\sim10^{-9}$, please see Tab.\ref{tab: PPN value} or \cite{Will2014}
for more details. Generally, the coordinate velocity $w$ of the PPN
coordinate system relative to the mean rest-frame of the universe
is generally believed to be small, that $w\sim\mathcal{O}(\epsilon)$
\cite{Will1972,Will1993,Will2014}. Therefore, the gradients produced
by the preferred-frame and the preferred-location effects in orbiting
SGGs will be smaller than $10^{-21}s^{-2},$ which is too small to
be seen by the present day SGGs and will be ignored in this work.
The above metric can then be cast into a rather simple form
\begin{eqnarray}
 &  & g_{\mu\nu}=\nonumber \\
 &  & \left(\begin{array}{cccc}
-1+\frac{2M}{r}-\frac{2\beta M^{2}}{r^{2}} & (\frac{\Delta x^{2}}{r^{3}}+\frac{3\chi x^{1}x^{3}}{2r^{4}})J & (-\frac{\Delta x^{1}}{r^{3}}+\frac{3\chi x^{2}x^{3}}{2r^{4}})J & -\frac{\chi[(x^{1})^{2}+(x^{2})^{2}-2(x^{3})^{2}]}{2r^{4}}J\\
\\
(\frac{\Delta x^{2}}{r^{3}}+\frac{3\chi x^{1}x^{3}}{2r^{4}})J & 1+\frac{2\gamma M}{r} & 0 & 0\\
\\
(-\frac{\Delta x^{1}}{r^{3}}+\frac{3\chi x^{2}x^{3}}{2r^{4}})J & 0 & 1+\frac{2\gamma M}{r} & 0\\
\\
-\frac{\chi[(x^{1})^{2}+(x^{2})^{2}-2(x^{3})^{2}]}{2r^{4}}J & 0 & 0 & 1+\frac{2\gamma M}{r}
\end{array}\right),\nonumber \\
\label{eq:metric}
\end{eqnarray}
where $r=\sqrt{\delta_{ij}x^{i}x^{j}}$, $\Delta=1+\gamma+\frac{1}{4}\alpha_{1}$,
and $M,\ \mathbf{J}$ are the asymptotically measured total mass and
angular momentum of Earth
\begin{eqnarray*}
M & = & \int\rho[1+(\gamma+1)v^{2}+(3\gamma-2\beta+1)U+\frac{\Pi}{\rho}+3\gamma\frac{p}{\rho}]d^{3}x,\\
\mathbf{J} & = & \int\rho(\mathbf{x}\times\mathbf{v})d^{3}x.
\end{eqnarray*}
For a satellite orbiting Earth with velocity $v,$ one has the basic
order relations
\begin{eqnarray}
v^{2}\sim\frac{M}{r}\sim\mathcal{O}(\epsilon^{2}),\quad\: v^{4}\sim\frac{M^{2}}{r^{2}}\sim\frac{Jv}{r^{2}}\sim\mathcal{O}(\epsilon^{4}).\label{eq:MJorder}
\end{eqnarray}

\subsection{Gravity gradient signals in non-dynamical Chern-Simons Gravity\label{sub:Gravity-gradient-signals}}

\begin{figure}
\center\includegraphics[scale=0.7]{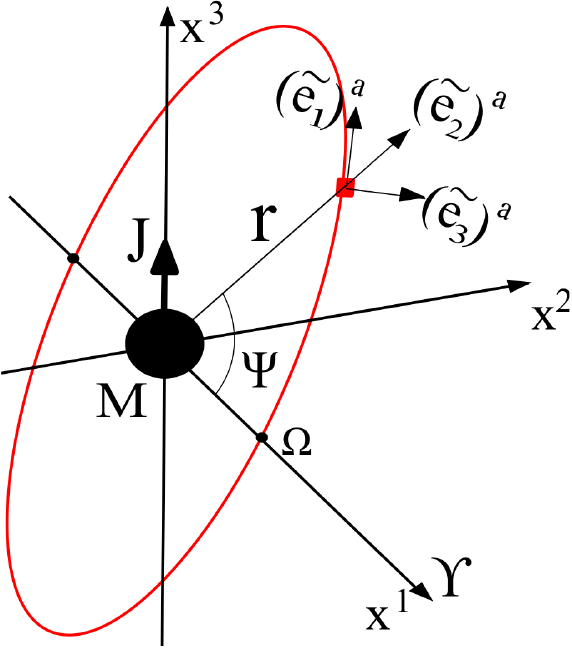}
\protect\caption{The orbits under consideration are circular orbits with the 
longitude
of ascending node $\Omega=0$. As illustrated in the figure, the local
tetrad carried by the S/C is defined as follows, $(\tilde{\mathbf{e}}_{1})^{a}$
is set along the direction of motion of the S/C, $(\tilde{\mathbf{e}}_{2})^{a}$
along the radial direction $\hat{\mathbf{r}}$ and $(\tilde{\mathbf{e}}_{3})^{a}=(\tilde{\mathbf{e}}_{1})^{a}\times(\tilde{\mathbf{e}}_{2})^{a}$.}
\label{fig:coor}
\end{figure}

For clearness, we set the spacecraft (S/C) that carrying the multi-axes
SGG to follow a circular orbit
\begin{equation}
x^{1}=a\cos\Psi,\ x^{2}=a\cos i\sin\Psi,\ x^{3}=a\sin i\sin\Psi,\label{eq:orbit}
\end{equation}
where the longitude of ascending node $\Omega$, the eccentricity
$e$ are set to be zero, $i$ and $a$ denote the inclination and
semi-major axis, see figure.\ref{fig:coor}. $\Psi=\omega\tau$ is
the orbital phase, and $\tau$ is the proper time measured by the
on-board clock.
We calculate the gravitational gradients measured along the above
orbit for the Earth pointing option of S/C attitude. The local Earth
pointing frame attached to the S/C is determined by the following
local tetrad, that we set $(\tilde{\mathbf{e}}_{1})^{a}$ along the
direction of motion of the S/C, $(\tilde{\mathbf{e}}_{2})^{a}$ along
the radial direction $\hat{\mathbf{r}}$, $(\tilde{\mathbf{e}}_{3})^{a}=(\tilde{\mathbf{e}}_{1})^{a}\times(\tilde{\mathbf{e}}_{2})^{a}$
and $(\tilde{\mathbf{e}}_{0})^{a}=Z^{a}$ the 4-velocity of the S/C,
see again figure.\ref{fig:coor}. Here we do not employ the initial
guided option of the S/C attitude since the gradient signal from the
gravitomagnetic sector will then have twice the orbital frequency
\cite{Mashhoon1989} and can hardly be separated from the far more
larger signal produced by Earth $J_{2}$ component at the same frequency
band. Also, as one will see, with the help of Earth pointing attitude
option, the gradient signal of the CS gravity can be separated from
that of the semi-conservative metric theory.

The geodesic deviation equation reads
\begin{equation}
Z^{b}\nabla_{b}Z^{c}\nabla_{c}\xi^{a}+R_{bcd}^{\ \ \ \ a}Z^{b}Z^{d}\xi^{c}=0,\label{eq:deviation}
\end{equation}
where $\xi^{a}$ is the connection vector determined by the axis of
the SGG. The gravitational gradient tensor or the tidal tensor is
then defined as
\begin{equation}
\tilde{K}_{\nu}^{\ \mu}=R_{\rho\beta\lambda}^{\ \ \ \ \alpha}Z^{\rho}Z^{\lambda}\mathbf{\mathbf{e}_{\nu}^{\ \beta}\mathbf{\underline{e}_{\alpha}^{\ \mu}}},\label{eq:gradient}
\end{equation}
here we use $\tilde{}$ to mark tensor components under the local
tetrad $(\tilde{e}_{\mu})^{a}$ to distinguish the components under
the original PN basis $(\frac{\partial}{\partial x^{\mu}})^{a}$,
and $\mathbf{e}_{\mu}^{\ \alpha}$ denotes the transformation matrix
from local frame to the Earth centered PN system, and $\mathbf{\underline{e}_{\alpha}^{\ \mu}}$
denotes the inverse. From dimensional analysis, we have
\[
\tilde{K}_{\nu}^{\ \mu}=\frac{1}{a^{2}}[\underbrace{\mathcal{O}(\frac{M}{a})+\mathcal{O}(v^{2})}_{Newtonian\ terms}+\underbrace{\mathcal{O}(\frac{M^{2}}{a^{2}})+\mathcal{O}(\frac{Mv^{2}}{a})+\mathcal{O}(\frac{Jv}{a^{2}})}_{Post-Newtonian\ terms}...],
\]
which means that to calculate the 1PN tidal components, that the $\frac{1}{a^{2}}\mathcal{O}(\epsilon^{4})$
terms, only the Keplerian orbits of the S/C with fixed orbital elements
are needed. To evaluate the Newtonian tidal components, the 1PN correction
to Keplerian orbits is then needed. However, since the Newtonian components
depends only on the semi-major $a$ and the unit position vector $\hat{\mathbf{r}}$,
one can then write $\hat{\mathbf{r}}$ in terms of the instantaneous
orbit elements evaluated at the time of observation to work out the
Newtonian tidal components. The full solutions of the 1PN orbits in
CS modified gravity will be the subjects of a separated publication.

The 4-velocity of the S/C reads
\begin{equation}
Z^{a}=\frac{dt}{d\tau}(\frac{\partial}{\partial t})^{a}+a\omega[-\sin\Psi(\frac{\partial}{\partial x^{1}})^{a}+\cos i\cos\Psi(\frac{\partial}{\partial x^{2}})^{a}+\sin i\cos\Psi(\frac{\partial}{\partial x^{3}})^{a}].\label{eq:Z}
\end{equation}
The ratio $\frac{dt}{d\tau}$ can be derived from the line element
$d\tau^{2}=-g_{\mu\nu}dx^{\mu}dx^{\nu}$ along the circular orbit,
to the required order we have
\begin{eqnarray}
\frac{dt}{d\tau} & = & 1+\frac{r^{2}\omega^{2}}{2}+\frac{M}{a}+\mathcal{O}(\epsilon^{4}).\label{eq:time}
\end{eqnarray}
For the tetrad attached to the reference mass defined in the last
subsection, we first set $(\tilde{\mathbf{e}}_{0})^{a}=Z^{a}$, and
following the Gram-Schmidt process the three spacial tetrad can be
worked out to the required order as
\begin{eqnarray}
(\tilde{\mathbf{e}}_{1})^{a} & = & a\omega(\frac{\partial}{\partial t})^{a}+(1+\frac{a^{2}\omega^{2}}{2}-\frac{\gamma M}{a})[-\sin\Psi(\frac{\partial}{\partial x^{1}})^{a}+\cos i\cos\Psi(\frac{\partial}{\partial x^{2}})^{a}\nonumber \\
 &  & +\sin i\cos\Psi(\frac{\partial}{\partial x^{3}})^{a}],\label{eq:e1}\\
(\tilde{\mathbf{e}}_{2})^{a} & = & \left(1-\frac{\gamma M}{a}\right)[\cos\Psi(\frac{\partial}{\partial x^{1}})^{a}+\cos i\sin\Psi(\frac{\partial}{\partial x^{2}})^{a}+\sin i\sin\Psi(\frac{\partial}{\partial x^{3}})^{a}],\nonumber \\
\label{eq:e2}\\
(\tilde{\mathbf{e}}_{3})^{a} & = & \left(1-\frac{\gamma M}{a}\right)[\cos i(\frac{\partial}{\partial x^{3}})^{a}-\sin i(\frac{\partial}{\partial x^{2}})^{a}].\label{eq:e3}
\end{eqnarray}
The transformation matrices $\mathbf{e}_{\mu}^{\ \alpha}$ and $\mathbf{\underline{e}_{\alpha}^{\ \mu}}$
are worked out to 1PN level, please see Eq.(\ref{eq:e}) and Eq.(\ref{eq:ie})
in Appendix.\ref{sec:tidal matrix}.

With Eq.(\ref{eq:gradient}), Eq.(\ref{eq:Z}), and Eq.(\ref{eq:e})-(\ref{eq:ie}),
the explicit forms of the 3-dimensional 1PN tidal matrix in the Earth
pointing local frame can be worked out. For clarity, we move the explicit
expressions in Appendix.\ref{sec:tidal matrix}, please see Eq.(\ref{eq:TN})-(\ref{eq:Tw})
for details. These tidal matrices agree with the results in \cite{Paik1988,Paik1989,Mashhoon1989}
that, in measuring the off-diagonal components of $\tilde{K}_{ij}$,
the gradient signals from the gravitomagnetic sector can be separated
from those produced by the Newtonian and 1PN gravitoelectric sectors.
Now, the most interesting result turns out to be that the gradient
signature in non-dynamical CS theory can be separated from the gravitomagnetic
tidal signals of the standard parity-conserving metric theories in
$\tilde{K}_{12}$ and $\tilde{K}_{23}$ components, see Eq.(\ref{eq:TGM})
and Eq.(\ref{eq:TCS}).

\begin{figure}
\center\includegraphics[scale=0.4]{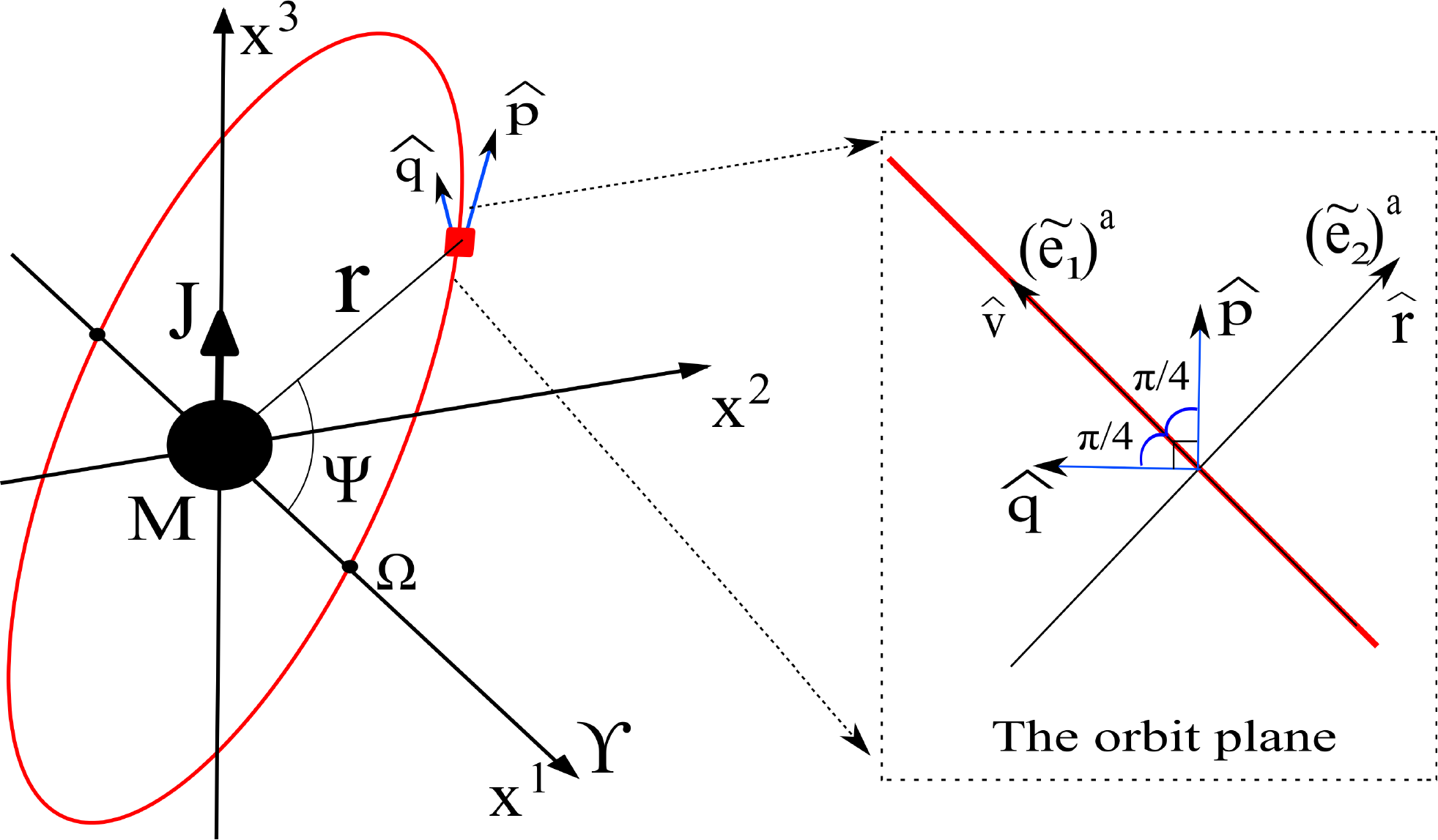}

\protect\caption{The configuration of the two-axes SGG along the circular orbit. The
two perpendicular axes $\hat{\mathbf{p}}$ and $\hat{\mathbf{q}}$,
Eq.(\ref{eq:pq}), lie within the orbital plane, and they both form
a angle of $\pi/4$ with the tangent vector of the orbit. }
\label{fig:2SGG}
\end{figure}

With a proper orientation of a two-axes diagonal-component SGG, one
can read out these off-diagonal components produced by the CS modification.
Denote $\hat{\mathbf{p}}$ and $\hat{\mathbf{q}}$ as the orthogonal
normalized 3-vectors of the two axes of the on-board SGG, and in the
S/C local frame we set
\begin{eqnarray}
\hat{\mathbf{p}} & = & \{\frac{1}{\sqrt{2}},\ \frac{1}{\sqrt{2}},\ 0\},\ \ \ \ \hat{\mathbf{q}}=\{\frac{1}{\sqrt{2}},\ -\frac{1}{\sqrt{2}},\ 0\},\label{eq:pq}
\end{eqnarray}
please see Fig.\ref{fig:2SGG}.
According to the tidal matrices Eq.(\ref{eq:TN})-Eq.(\ref{eq:Tw}),
the sum and difference of the SGG read outs $\tilde{K}_{\hat{\mathbf{p}}\hat{\mathbf{p}}}$
and $\tilde{K}_{\hat{\mathbf{q}}\hat{\mathbf{q}}}$ along these two
axes can be written down as
\begin{eqnarray}
\tilde{K}^{+} & \equiv & \tilde{K}_{\hat{\mathbf{p}}\hat{\mathbf{p}}}+\tilde{K}_{\hat{\mathbf{q}}\hat{\mathbf{q}}}\nonumber \\
 & = & -\frac{M}{a^{3}}+\frac{(4\beta+2\gamma-3)M^{2}}{a^{4}}-\frac{(\gamma+2)\omega^{2}M}{a}\nonumber \\
 &  & +\frac{3J\omega\Delta\cos i}{a^{3}}-2\omega_{0}^{2}-\frac{3J\chi\omega\sin i\cos\Psi}{2a^{3}},\label{eq:K+}\\
\tilde{K}^{-} & \equiv & \tilde{K}_{\hat{\mathbf{p}}\hat{\mathbf{p}}}-\tilde{K}_{\hat{\mathbf{q}}\hat{\mathbf{q}}}=-\frac{J\chi\omega\sin i\sin\Psi}{2a^{3}},\label{eq:K-}
\end{eqnarray}
here $\omega_{0}$ denote the angular velocity of the rotating local
frame relative to the parallel transported frame. Please see Appendix.\ref{sec:SGG readout}
for the explicit form of the readouts along $\hat{\mathbf{p}}$ and
$\hat{\mathbf{q}}$. Therefore, for this simple two-axes SGG settings,
the gradient signal produced by the CS modification can be isolated,
especially in $\tilde{K}^{-}$, from those produced by the gravitomagnetic
sector of the standard parity-conserving metric theory.

\begin{figure}
\center\includegraphics[scale=0.5]{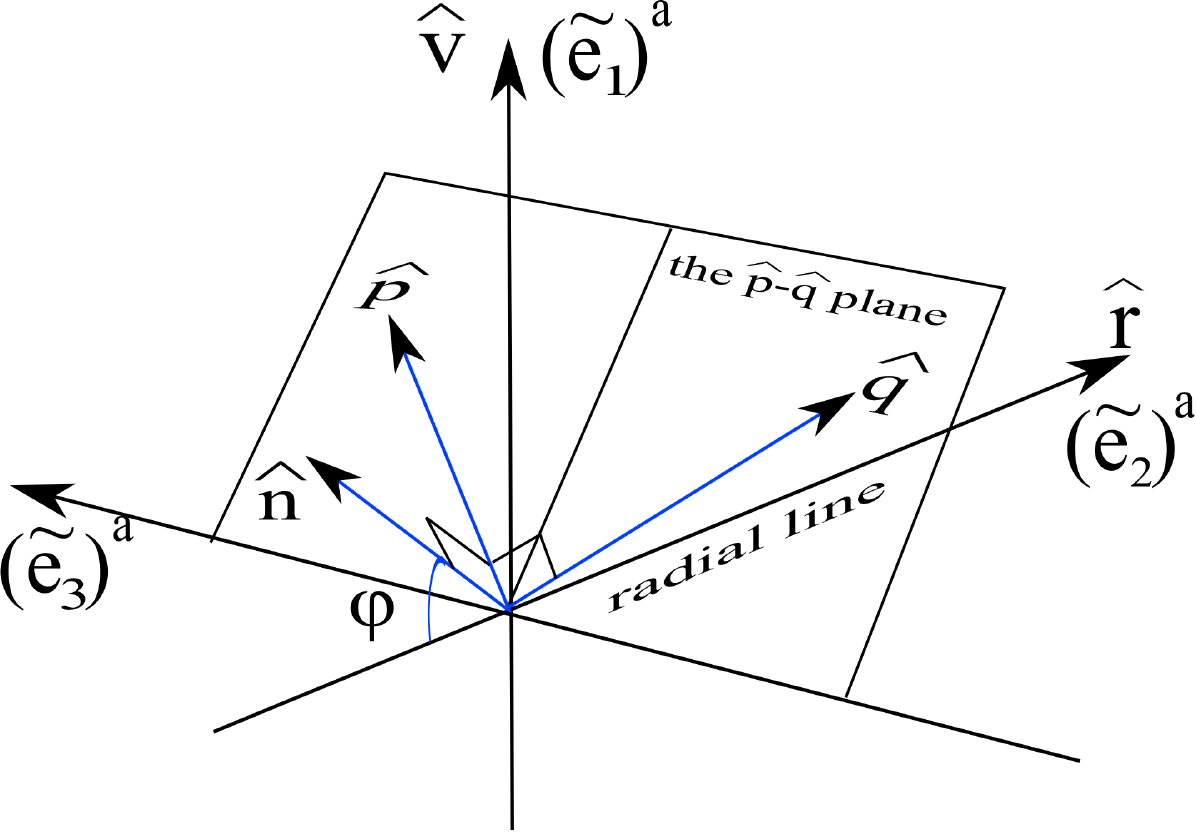}

\protect\caption{The configurations of three-axes SGG along the circular orbit. As
in Eq.(\ref{eq:n3})-(\ref{eq:q3}), The two perpendicular axes $\hat{\mathbf{p}}$
and $\hat{\mathbf{q}}$ are symmetric to the $\hat{\mathbf{r}}-\hat{\mathbf{v}}$
plane, and $\hat{\mathbf{n}}$ is orthogonal to the $\hat{\mathbf{p}}-\hat{\mathbf{q}}$
plane. The angle between $\hat{\mathbf{n}}$ and $-\hat{\mathbf{r}}$
is denoted as $\phi$.}
\label{fig:3SGG}
\end{figure}

For the more reliable three-axes SGG options studied in\cite{Paik1988,Mashhoon1989,Paik1989},
we now set the three orthogonal SGG axes in the S/C local frame as
\begin{eqnarray}
\hat{\mathbf{n}} & = & \{\sin\phi,\ -\cos\phi,\ 0\},\label{eq:n3}\\
\hat{\mathbf{p}} & = & \frac{1}{\sqrt{2}}\{\cos\phi,\ \sin\phi,\ -1\},\label{eq:p3}\\
\hat{\mathbf{q}} & = & \frac{1}{\sqrt{2}}\{\cos\phi,\ \sin\phi,\ 1\},\label{eq:q3}
\end{eqnarray}
where $\phi$ is the angle between $\hat{\mathbf{n}}$ and $-\hat{\mathbf{r}}$.
please see Fig.\ref{fig:3SGG}.
As before, one can use the combinations of $\tilde{K}_{\hat{\mathbf{n}}\hat{\mathbf{n}}}$,
$\tilde{K}_{\hat{\mathbf{p}}\hat{\mathbf{p}}}$ and $\tilde{K}_{\hat{\mathbf{q}}\hat{\mathbf{q}}}$
to single out the interested signals
\begin{eqnarray}
\tilde{K}^{+} & \equiv & \tilde{K}_{\hat{\mathbf{p}}\hat{\mathbf{p}}}+\tilde{K}_{\hat{\mathbf{q}}\hat{\mathbf{q}}}\nonumber \\
 & = & \frac{M(3\cos2\phi+1)}{2a^{3}}-\omega_{0}^{2}-\frac{((4\gamma-1)+(8\beta+8\gamma-7)\cos2\phi)M^{2}}{4a^{4}}\nonumber \\
 &  & +\frac{((\gamma+2)\cos2\phi+3\gamma)M\omega^{2}}{4a}-\frac{3J\omega\Delta\cos i\cos^{2}\phi}{a^{3}}\nonumber \\
 &  & +\frac{J\omega(\chi\sin i(3\cos2\phi-1)\cos\Psi-2\chi\sin i\sin\phi\cos\phi\sin\Psi)}{4a^{3}},\label{eq:3K+}
\end{eqnarray}
\begin{eqnarray}
\tilde{K}^{-} & \equiv & \tilde{K}_{\hat{\mathbf{p}}\hat{\mathbf{p}}}-\tilde{K}_{\hat{\mathbf{q}}\hat{\mathbf{q}}}\nonumber \\
 & = & \frac{3J\omega\Delta\sin i(\cos\phi\cos\Psi-3\sin\phi\sin\Psi)}{a^{3}}+\frac{J\omega\chi\cos i\cos\phi}{2a^{3}},\label{eq:3K-}
\end{eqnarray}
\begin{eqnarray}
\tilde{K}^{\oplus} & \equiv & \tilde{K}_{\hat{\mathbf{p}}\hat{\mathbf{p}}}+\tilde{K}_{\hat{\mathbf{q}}\hat{\mathbf{q}}}+\tilde{K}_{\hat{\mathbf{n}}\hat{\mathbf{n}}}\nonumber \\
 & = & -\frac{J\chi\omega\sin i\cos\Psi}{a^{3}}+\frac{M^{2}(2\beta-\gamma-1)}{a^{4}}+\frac{(\gamma-1)M\omega^{2}}{a}-2\omega_{0}^{2}.\label{eq:3K++}
\end{eqnarray}
Please see Appendix.\ref{sec:SGG readout} for the explicit form of
the readouts along $\hat{\mathbf{n}}$, $\hat{\mathbf{p}}$ and $\hat{\mathbf{q}}$.
Therefore, without any change of the optimized three-axes configurations
studied in \cite{Paik1988,Paik1989,Mashhoon1989}, the above results
turns out to be quite promising. The only periodic signal in the difference
$\tilde{K}^{-}$ is from the parity-conserving metric theories, which
can be distinguished from the DC signal produced by the CS modification
and then be used to test the gravitomagnetic sector of such theories
as suggested in\cite{Paik1988,Paik1989,Mashhoon1989,Paik2008}. For
the key result here, the only periodic signal $-\frac{J\chi\omega\sin i\cos\Psi}{a^{3}}$
in the total sums $\tilde{K}^{\oplus}$ (\emph{or the trace of the
total tidal matrix}) comes from the parity-violating CS modification,
which can be isolated from the rest DC terms contained in $\tilde{K}^{\oplus}$.

\section{The measurements of the Chern-Simons gradient
signal\label{sec:measurements}}

Recovering the SI units, the CS gradient signal to be measured is
proportional to $\frac{GJ\chi\omega\sin i\cos\Psi}{c^{2}a^{3}}$.
To estimate the magnitudes of the signals and the accuracy requirements
for our experiment, we here assume a polar circular orbit with specific
altitude as $650km$, that $a=7020km$, just like in the GP-B mission.
For the three-axes SGG, the magnitude of the CS gradient signal is
about $1.5\times10^{-17}\chi s^{-2}$ with frequency about $0.17mHz$.
As reported in \cite{Moody2002}, the fully developed SGG at the University
of Maryland with mechanically suspended test masses has the sensitivity
of $10^{-11}s^{-2}/Hz^{\frac{1}{2}}$ in the signal frequency band.
For one year lifetime missions, that $t\sim\pi\times10^{7}s$, the
CS parameter $\chi$ can, in principle, be constrained to be smaller
than $10^{2}$. From Eq.(\ref{eq:mass}), recovering the SI units,
we have
\[
M_{CS}=\frac{4\hbar c}{\chi a}.
\]
Therefore, for this conservative option of the SGG sensitivity, the
CS mass scale can be constrained to be larger than $1.1\times10^{-15}eV.$
This is of course a rather weak bound on the CS scalar $\theta$ or
the mass scale of the theory. While, as mentioned before, the high
precision SGG with performance level about $10^{-14}s^{-2}Hz^{-\frac{1}{2}}$
in the band $10^{-4}Hz\sim10^{-1}Hz$ has already been developed at
the University of Maryland \cite{Moody2010,Shirron2010}, and even
for the performance level of $10^{-15}s^{-2}Hz^{-\frac{1}{2}}$ at
the signal frequency band is within the capability of the SGGs under
development \cite{Paik2006}. To be optimistic, with the likely 4
orders of magnitude improvement in the SGG performance level in the
future, the CS mass scale $M_{CS}$ could be constrained to be lager
than $1.1\times10^{-11}eV$ for one year lifetime missions.

The signal to be measured will be polluted by various noises, especially
those from the misalignment between axes and Earth gravity multiples.
One important point is that the pointing error of the axes will not
affect our measurement, since the combination $\tilde{K}^{\oplus}$
is the trace of the tidal matrix that is invariant under axes rotations.
For our measurement is similar to that of the gravitomagnetic signal
in the combination of $\tilde{K}^{-}$ as discussed in \cite{Paik1988,Mashhoon1989,Paik2008},
one can then consult \cite{Paik2008,Li2014} for the analysis of the
noises from misalignments and Earth multiples. As pointed out in \cite{Paik2008},
for Earth pointing orientation, the couplings between the alignment
error with Newtonian and 1PN gravitoelectric gradients will not contribute
since the alignment does not modulate the these gradients in such
orientation. While, for the gravitomagnetic gradient from the parity-conserving
theories, its couplings with the alignment error will enter into the
trace signal $\tilde{K}^{\oplus}$. But, to reach the constraint that
$\chi<10^{-2}$ mentioned above one only need the misalignments to
be measured to $10^{-3}rad$ in one year, which is rather an trivial
task in our experiment. Thus, the test of the CS modified gravity
will not suffer from the difficulties of meeting the stringent alignments
and pointing requirements as in the measurements of the gravitomagnetic
signal of standard parity-conserving theories. The detailed analysis concerning
all these error sources will be the subjects of a future work.

At last, in the measurements of gravitomagnetic effects from parity-conserving
theories, secular signals with increasing magnitude exist \cite{Mashhoon1982,Theiss1985}
due to the frame-dragging effect that causes the SGG axes to tilt
relative to Earth with angular velocity $\sim\frac{J}{a^{3}}t$ \cite{Schiff1960,Paik2008}.
These secular signals make the detections of the gravitomagnetic effect
much easier. The non-dynamical CS modified gravity differs with the
parity-conserving metric theories only in the gravitomagnetic sector.
Therefore, it is natural to ask whether or not similar secular gradient
signals exist for the non-dynamical CS modified gravity, since the
axes (or gyros) may tilt differently along the orbit due to a different
gravitomagnetic torque. The answers to this problem require the studies
of the 1PN geodesic motions and spin precession in the CS modified
gravity, which themselves form an interesting subject and will left
in our future works.

To conclude here, we studied in this paper the theoretical principles
for testing non-dynamical CS modified gravity with SGGs in space.
For the two-axes and three-axes diagonal-component SGGs, we worked
out the characteristic signals from the CS modification in the combinations
of the measurements of different axes, which can be clearly isolated
in these combined data. Furthermore, for the three-axes SGG option,
the precision tests of the gravitomagnetic effect in parity-conserving
metric theories\cite{Paik1988,Paik1989,Mashhoon1989} and the test
of the effects from CS modification can be carried out independently
in the same time. High precision multi-axes SGGs are powerful tools
for relativistic experiments, especially that carried in space, and
it is promising to add the ability of testing the string theory inspired
CS modified gravity to the space based experiments with orbiting SGGs.
\begin{acknowledgements}
This work was supported by the NSFC grands No. 41204051. and No. 11305255.
\end{acknowledgements}

\appendix

\section{The standard PPN metric\label{sec:The-standard-PPN}}

The standard PPN metric has the form \cite{Will2014}

\begin{eqnarray*}
g_{00} & = & -1+2U-2\beta U^{2}-2\xi\Phi_{W}+(2\gamma+2+\alpha_{3}+\zeta_{1}-2\xi)\Phi_{1}\\
 &  & +2(3\gamma-2\beta+1+\zeta_{2}+\xi)\Phi_{2}+2(1+\zeta_{3})\Phi_{3}+2(3\gamma+3\zeta_{4}-2\xi)\Phi_{4}\\
 &  & -(\zeta_{1}-2\xi)\mathcal{A}-(\alpha_{1}-\alpha_{2}-\alpha_{3})w^{2}U-\alpha_{2}w^{i}w^{j}U_{ij}+(2\alpha_{3}-\alpha_{1})w^{i}V_{i}+\mathcal{O}(\epsilon^{6}),\\
g_{0i} & = & -\frac{1}{2}(4\gamma+3+\alpha_{1}-\alpha_{2}+\zeta_{1}-2\xi)V_{i}-\frac{1}{2}(1+\alpha_{2}-\zeta_{1}+2\xi)W_{i}\\
 &  & -\frac{1}{2}(\alpha_{1}-2\alpha_{2})w^{i}U-\alpha_{2}w^{j}U_{ij}+\mathcal{O}(\epsilon^{5}),\\
g_{ij} & = & (1+2\gamma U)\delta_{ij}+\mathcal{O}(\epsilon^{4}),
\end{eqnarray*}
where the PN potentials read
\begin{eqnarray*}
U & = & \int\frac{\rho'}{|\mathbf{x}-\mathbf{x'}|}d^{3}x',\ \ \ \ \Phi_{1}=\int\frac{\rho'v'^{2}}{|\mathbf{x}-\mathbf{x'}|}d^{3}x',\\
\Phi_{2} & = & \int\frac{\rho'U'}{|\mathbf{x}-\mathbf{x'}|}d^{3}x',\ \ \ \ \Phi_{3}=\int\frac{\rho'\Pi'}{|\mathbf{x}-\mathbf{x'}|}d^{3}x',\\
\Phi_{4} & = & \int\frac{p'}{|\mathbf{x}-\mathbf{x'}|}d^{3}x',\ \ \ \ V_{i}=\int\frac{\rho'v'^{i}}{|\mathbf{x}-\mathbf{x'}|}d^{3}x',\\
W_{i} & = & \int\frac{\rho'[\mathbf{v}'\cdot(\mathbf{x}-\mathbf{x'})](x^{i}-x'^{i})}{|\mathbf{x}-\mathbf{x'}|^{3}}d^{3}x',\\
U_{ij} & = & \int\frac{\rho'(x^{i}-x'^{i})(x^{j}-x'^{j})}{|\mathbf{x}-\mathbf{x'}|^{3}}d^{3}x',\\
\mathcal{A} & = & \int\frac{\rho'[\mathbf{v}'\cdot(\mathbf{x}-\mathbf{x'})]^{2}}{|\mathbf{x}-\mathbf{x'}|^{3}}d^{3}x',\\
\Phi_{W} & = & \int\frac{\rho'\rho''(\mathbf{x}-\mathbf{x}')}{|\mathbf{x}-\mathbf{x'}|^{3}}\cdot(\frac{\mathbf{x'}-\mathbf{x}''}{|\mathbf{x'}-\mathbf{x}''|}-\frac{\mathbf{x}-\mathbf{x}''}{|\mathbf{x}-\mathbf{x}''|})d^{3}x'd^{3}x''.
\end{eqnarray*}
The matter variables are the rest mass density $\rho$, pressure $p$,
coordinate velocity of the matter field $v^{i}$, internal energy
per unit mass $\Pi$ and the coordinate velocity of the PPN coordinate
system relative to the mean rest-frame of the universe $w^{i}$. The
PN orders read
\[
v\sim\mathcal{O}(\epsilon),\ \ \ \ v^{2}\sim U\sim\Pi\sim\frac{p}{\rho}\sim\mathcal{O}(\epsilon^{2}).
\]

The standard PN parameters $\{\gamma,\ \beta,\ \xi\ ,\alpha_{1},\ \alpha_{2},\ \alpha_{3},\ \zeta_{1},\ \zeta_{2},\ \zeta_{3},\ \zeta_{4}\}$
have the following meanings. The parameters $\gamma$ and $\beta$
are the usual Eddington--Robertson--Schiff parameters used to describe
the ``classical'' tests of GR and are in some sense the most important
ones. For GR $\gamma=\beta=1$ are the only non-vanishing parameters.
The parameter $\xi$ measures the preferred-location effects, $\{\alpha_{1},\ \alpha_{2},\ \alpha_{3}\}$
measure the preferred-frame effects and$\{\alpha_{3},\ \zeta_{1},\ \zeta_{2},\ \zeta_{3},\ \zeta_{4}\}$
measure the violations of global conservation laws for total momentum.
The up-to-date values of these parameters are summarized in Tab.\ref{tab: PPN value}
\cite{Will2014}.
\begin{table}
\begin{tabular}{|c|c|c|}
\hline
\textbf{Parameter}  & \textbf{Bound}  & \textbf{Experiment}\tabularnewline
\hline
\hline
$\gamma-1$  & $2.3\times10^{-5}$  & time delay in Cassini tracking\tabularnewline
\hline
 & $2\times10^{-4}$ & light deflection in VLBI\tabularnewline
\hline
$\beta-1$  & \multicolumn{1}{c|}{$8\times10^{-5}$ } & perihelion shift\tabularnewline
\hline
 & $2.3\times10^{-4}$ & Nordtvedt effect\tabularnewline
\hline
$\xi$  & $10^{-9}$  & spin precession of millisecond pulsars\tabularnewline
\hline
$\alpha_{1}$  & $4\times10^{-5}$ & orbital polarization of PSR J1738+0333\tabularnewline
\hline
 & $10^{-4}$  & Lunar laser ranging\tabularnewline
\hline
$\alpha_{2}$  & $2\times10^{-9}$  & spin precession of millisecond pulsars\tabularnewline
\hline
$\alpha_{3}$  & $4\times10^{-20}$  & pulsar spin down statistics\tabularnewline
\hline
$\zeta_{1}$  & 0.02  & combined PPN bounds\tabularnewline
\hline
$\zeta_{2}$  & $4\times10^{-5}$  & binary acceleration of PSR 1913+16\tabularnewline
\hline
$\zeta_{3}$  & $10^{-8}$  & Lunar acceleration\tabularnewline
\hline
$\zeta_{4}$  & --- & not independent\tabularnewline
\hline
\end{tabular}\protect\caption{Current values of PPN parameters.}
\label{tab: PPN value}
\end{table}

\section{The transformation matrices and tidal matrices\label{sec:tidal matrix}}

From Eq.(\ref{eq:Z})-Eq.(\ref{eq:e3}), the transformation matrices
between the local frame and the Earth centered PN system up to 1PN
level read
\begin{eqnarray}
 &  & \mathbf{e=}\nonumber \\
 &  & \left(\begin{array}{cccc}
1+\frac{a^{2}\omega^{2}}{2}+\frac{M}{a} & -a\omega\sin\Psi & a\omega\cos i\cos\Psi & a\omega\sin i\cos\Psi\\
\\
a\omega & -(1+\frac{a^{2}\omega^{2}}{2}-\frac{\gamma M}{a})\sin\Psi & (1+\frac{a^{2}\omega^{2}}{2}-\frac{\gamma M}{a})\cos i\cos\Psi & (1+\frac{a^{2}\omega^{2}}{2}-\frac{\gamma M}{a})\sin i\sin\Psi\\
\\
0 & \left(1-\frac{\gamma M}{a}\right)\cos\Psi & \left(1-\frac{\gamma M}{a}\right)\cos i\sin\Psi & \left(1-\frac{\gamma M}{a}\right)\sin i\sin\Psi\\
\\
0 & 0 & -\left(1-\frac{\gamma M}{a}\right)\sin i & \left(1-\frac{\gamma M}{a}\right)\cos i
\end{array}\right),\nonumber \\
\label{eq:e}
\end{eqnarray}
\begin{eqnarray}
 &  & \underline{\mathbf{e}}=\nonumber \\
 &  & \left(\begin{array}{cccc}
1+\frac{a^{2}\omega^{2}}{2}-\frac{M}{a} & -a\omega & 0 & 0\\
\\
a\omega\sin\Psi & -(1+\frac{a^{2}\omega^{2}}{2}+\frac{\gamma M}{a})\sin\Psi & (1+\frac{\gamma M}{a})\cos\Psi & 0\\
\\
-a\omega\cos i\cos\Psi & (1+\frac{a^{2}\omega^{2}}{2}+\frac{\gamma M}{a})\cos i\cos\Psi & (1+\frac{\gamma M}{a})\cos i\sin\Psi & -(1+\frac{\gamma M}{a})\sin i\\
\\
-a\omega\sin i\cos\Psi & (1+\frac{a^{2}\omega^{2}}{2}+\frac{\gamma M}{a})\sin i\cos\Psi & (1+\frac{\gamma M}{a})\sin i\sin\Psi & (1+\frac{\gamma M}{a})\cos i
\end{array}\right).\nonumber \\
\label{eq:ie}
\end{eqnarray}

From Eq.(\ref{eq:gradient}), Eq.(\ref{eq:Z}), and Eq.(\ref{eq:e})-(\ref{eq:ie}),
the 3 dimensional tidal matrix $\tilde{K}_{ij}\text{\ensuremath{\equiv}}\tilde{K}_{ij}^{N}+\tilde{K}_{ij}^{GE}+\tilde{K}_{ij}^{GM}+\tilde{K}_{ij}^{CS}$
along the circular orbit in the Earth pointing local frame can be
worked out as
\begin{equation}
\tilde{K}^{N}=\frac{M}{a^{3}}\left(\begin{array}{ccc}
1 & 0 & 0\\
0 & -2 & 0\\
0 & 0 & 1
\end{array}\right),\label{eq:TN}
\end{equation}
\begin{equation}
\tilde{K}^{GE}=\frac{M}{a^{3}}\left(\begin{array}{ccc}
-\frac{(2\beta+3\gamma-2)M}{a} & 0 & 0\\
\\
0 & \frac{(6\beta+5\gamma-5)M}{a}-(\gamma+2)a^{2}\omega^{2} & 0\\
\\
0 & 0 & \frac{(-2\beta-3\gamma+2)M}{a}+(2\gamma+1)\omega^{2}a^{2}
\end{array}\right),\label{eq:TGE}
\end{equation}

\begin{equation}
\tilde{K}^{GM}=\frac{J\omega}{a^{3}}\left(\begin{array}{ccc}
0 & 0 & -\frac{3}{2}\Delta\sin i\cos\Psi\\
\\
0 & 3\Delta\cos i & \frac{9}{2}\Delta\sin i\sin\Psi\\
\\
-\frac{3}{2}\Delta\sin i\cos\Psi & \frac{9}{2}\Delta\sin i\sin\Psi & -3\Delta\cos i
\end{array}\right),\label{eq:TGM}
\end{equation}
\begin{equation}
\tilde{K}^{CS}=\frac{J\omega}{a^{3}}\left(\begin{array}{ccc}
0 & -\frac{1}{4}\chi\sin i\sin\Psi & -\frac{1}{4}\chi\cos i\\
\\
-\frac{1}{4}\chi\sin i\sin\Psi & -\frac{3}{2}\chi\sin i\cos\Psi & 0\\
\\
-\frac{1}{4}\chi\cos i & 0 & \frac{1}{2}\chi\sin i\cos\Psi
\end{array}\right).\label{eq:TCS}
\end{equation}
Here, $\tilde{K}^{N}$, $\tilde{K}^{GE}$, $\tilde{K}^{GM}$ and $\tilde{K}^{CS}$
denote the gravitational tidal matrices from the Newtonian force,
the 1PN gravitoelectric force, the gravitomagnetic force and the contributions
from the CS modification. Since the Earth pointing frame is rotating
relative to parallel transported frames (Fermi shifted), we need to
include the tidal matrix from the centrifugal force produced by such
rotation
\begin{equation}
\tilde{K}^{\omega}=-\omega_{0}^{2}\begin{pmatrix}1\\
 & 1\\
 &  & 0
\end{pmatrix},\label{eq:Tw}
\end{equation}
where $\omega_{0}=\frac{d\Psi}{dt}+\frac{1}{a}\mathcal{O}(\epsilon^{3})$
denote the angular velocity of the rotation of the local frame.

\section{The explicit form of the gradients readouts\label{sec:SGG readout}}

For the two-axes SGG discussed in Subsection.\ref{sub:Gravity-gradient-signals},
see Eq.(\ref{eq:pq}) and Fig.\ref{fig:2SGG}, the readouts along
$\hat{\mathbf{p}}$ and $\hat{\mathbf{q}}$ are

\begin{eqnarray*}
\tilde{K}_{\hat{\mathbf{p}}\hat{\mathbf{p}}}^{\omega} & = & \tilde{K}_{\hat{\mathbf{q}}\hat{\mathbf{q}}}^{\omega}=-\omega_{0}^{2}\ \ \ \ \ \ \ \tilde{K}_{\hat{\mathbf{p}}\hat{\mathbf{p}}}^{N}=\tilde{K}_{\hat{\mathbf{q}}\hat{\mathbf{q}}}^{N}=-\frac{M}{2a^{3}},\\
\tilde{K}_{\hat{\mathbf{p}}\hat{\mathbf{p}}}^{GE} & = & \tilde{K}_{\hat{\mathbf{q}}\hat{\mathbf{q}}}^{GE}=\frac{(4\beta+2\gamma-3)M^{2}}{2a^{4}}-\frac{(\gamma+2)M\omega^{2}}{2a},\\
\tilde{K}_{\hat{\mathbf{p}}\hat{\mathbf{p}}}^{GM} & = & \tilde{K}_{\hat{\mathbf{q}}\hat{\mathbf{q}}}^{GM}=\frac{3\Delta J\omega\cos i}{2a^{3}},\\
\tilde{K}_{\hat{\mathbf{p}}\hat{\mathbf{p}}}^{CS} & = & -\frac{\chi J\omega\sin i(3\cos\Psi+\sin\Psi)}{4a^{3}},\ \ \ \ \ \tilde{K}_{\hat{\mathbf{q}}\hat{\mathbf{q}}}^{CS}=-\frac{\chi J\omega\sin i(3\cos\Psi-\sin\Psi)}{4a^{3}}.
\end{eqnarray*}
For the three-axes SGG in Subsection.\ref{sub:Gravity-gradient-signals},
see Eq.(\ref{eq:n3})-(\ref{eq:q3}) and Fig.(\ref{fig:3SGG}), the
readouts along $\hat{\mathbf{n}}$, $\hat{\mathbf{p}}$ and $\hat{\mathbf{q}}$
are
\begin{eqnarray*}
\tilde{K}_{\hat{\mathbf{n}}\hat{\mathbf{n}}}^{\omega} & = & -\omega_{0}^{2},\quad\tilde{K}_{\hat{\mathbf{p}}\hat{\mathbf{p}}}^{\omega}=\tilde{K}_{\hat{\mathbf{q}}\hat{\mathbf{q}}}^{\omega}=-\frac{\omega_{0}^{2}}{2},\\
\tilde{K}_{\hat{\mathbf{n}}\hat{\mathbf{n}}}^{N} & = & -\frac{M(3\cos2\phi+1)}{2a^{3}},\quad\tilde{K}_{\hat{\mathbf{p}}\hat{\mathbf{p}}}^{N}=\tilde{K}_{\hat{\mathbf{q}}\hat{\mathbf{q}}}^{N}=\frac{M(3\cos2\phi+1)}{4a^{3}},\\
\tilde{K}_{\hat{\mathbf{n}}\hat{\mathbf{n}}}^{GE} & = & \frac{\left((6\beta+5\gamma-5)\cos^{2}\phi-(2\beta+3\gamma-2)\sin^{2}\phi\right)M^{2}}{a^{4}}-\frac{(\gamma+2)\cos^{2}\phi M\omega^{2}}{a},\\
\tilde{K}_{\hat{\mathbf{p}}\hat{\mathbf{p}}}^{GE} & = & \tilde{K}_{\hat{\mathbf{q}}\hat{\mathbf{q}}}^{GE}=-\frac{((4\gamma-1)+(8\beta+8\gamma-7)\cos2\phi)M^{2}}{4a^{4}}+\frac{((\gamma+2)\cos2\phi+3\gamma)M\omega^{2}}{4a},\\
\tilde{K}_{\hat{\mathbf{n}}\hat{\mathbf{n}}}^{GM} & = & \frac{3\Delta J\omega\cos i\cos^{2}\phi}{a^{3}},\\
\tilde{K}_{\hat{\mathbf{p}}\hat{\mathbf{p}}}^{GM} & = & -\frac{3\Delta J\omega\left(\sin i(3\sin\phi\sin\Psi-\cos\phi\cos\Psi)+\cos i\cos^{2}\phi\right)}{2a^{3}},\\
\tilde{K}_{\hat{\mathbf{q}}\hat{\mathbf{q}}}^{GM} & = & -\frac{3\Delta J\omega\left(\sin i(-3\sin\phi\sin\Psi+\cos\phi\cos\Psi)+\cos i\cos^{2}\phi\right)}{2a^{3}},\\
\tilde{K}_{\hat{\mathbf{n}}\hat{\mathbf{n}}}^{CS} & = & \frac{J\chi\omega\sin i\cos\phi(\sin\phi\sin\Psi-3\cos\phi\cos\Psi)}{2a^{3}},\\
\tilde{K}_{\hat{\mathbf{p}}\hat{\mathbf{p}}}^{CS} & = & \frac{J\chi\omega\left(\sin i\left((1-3\sin^{2}\phi)\cos\Psi-\sin\phi\cos\phi\sin\Psi\right)+\cos i\cos\phi\right)}{4a^{3}},\\
\tilde{K}_{\hat{\mathbf{q}}\hat{\mathbf{q}}}^{CS} & = & \frac{J\chi\omega\left(\sin i\left((1-3\sin^{2}\phi)\cos\Psi-\sin\phi\cos\phi\sin\Psi\right)-\cos i\cos\phi\right)}{4a^{3}}.
\end{eqnarray*}

\bibliographystyle{spphys}
\addcontentsline{toc}{section}{\refname}\bibliography{Testing-Chern-Simons}

\end{document}